\documentclass[a4paper,11pt]{article}
\pdfoutput=1 

\usepackage{jheppub} 

\usepackage[T1]{fontenc} 
\usepackage{graphicx}
\usepackage{amsmath}
\usepackage{amssymb}
\usepackage{array}
\usepackage{slashed}
\usepackage{comment}
\usepackage{datetime}
\usepackage{soul}
\usepackage{pifont, subfigure}

\usepackage{float}
\usepackage{bm}
\usepackage{bbm,multicol}

\usepackage{subfigure}
\usepackage{dsfont}

\usepackage{marginnote}

\usepackage{multirow}

\def\h{h^0}
\def\H{H^0}
\def\A{A}
\newcommand{\sba}{\ensuremath{\sin(\beta-\alpha)}}
\newcommand{\cba}{\ensuremath{\cos(\beta-\alpha)}}
\newcommand{\hc}{H^{\pm}}
\newcommand{\w}{W^{\pm}}

\newcommand{\ifb}{\ensuremath{ \text{fb}^{-1}  }}
\newcommand{\cmark}{\ding{51}}%
\newcommand{\xmark}{\ding{55}}%

\title{Light Charged Higgs Bosons  to  $AW/HW$ via Top Decay}
\author{Felix Kling, Adarsh Pyarelal, Shufang Su}
\emailAdd{kling@email.arizona.edu, adarsh@email.arizona.edu, shufang@email.arizona.edu}
\affiliation{Department of Physics,
University of Arizona,
Tucson, AZ 85721, USA}

\abstract{
While current ATLAS and CMS measurements exclude a light charged Higgs ($m_{H^\pm}<160$ GeV)  for most of the parameter region in the context of the MSSM scenarios, these bounds are significantly weakened in the Type II 2HDM once the exotic decay channel into a lighter neutral Higgs,  $H^\pm \to AW/HW$,  is open. In this study, we examine the possibility of a light charged Higgs produced in top decay via single top or top pair production,   which is the most prominent production channel for a light charged Higgs at the LHC.   We consider the subsequent decay $H^\pm \rightarrow AW/HW$,  which can reach a sizable branching fraction  at low $\tan\beta$ once it is  kinematically permitted. With a detailed collider analysis,  we obtain exclusion and discovery bounds for the 14 TeV LHC assuming the existence of a 70 GeV neutral scalar.  Assuming ${\rm BR}(H^\pm \rightarrow AW/HW)=100\%$ and ${\rm BR}(A/H \rightarrow \tau\tau)=8.6\%$,  the 95\% exclusion limits on ${\rm BR}(t \rightarrow H^+ b)$ are about 0.2\% and 0.03\% for single top and top pair production respectively,  with an integrated luminosity of 300 ${\rm fb}^{-1}$.  The discovery  reaches  are about 3 times higher. In the context of the Type II 2HDM,   discovery is possible at both large $\tan\beta > 17$ for 155 GeV $< m_{H^\pm} <$ 165 GeV,  and small $\tan\beta < 6$ over the entire mass range.   Exclusion is possible in the entire $\tan\beta$ versus $m_{H^\pm}$ plane except for charged Higgs masses close to the top threshold. The exotic decay channel $H^\pm \to AW/HW$ is therefore complementary to the conventional $H^\pm \rightarrow \tau\nu$ channel.}

\begin{document}
\maketitle
\flushbottom
\newpage

\section{Introduction}
\label{sec:intro}

In July 2012,  both the ATLAS and the CMS collaborations announced the discovery of a new resonance with a mass of 126 GeV,  which is consistent with the predictions of the Standard Model (SM) Higgs boson \cite{Aad:2012tfa,  Chatrchyan:2012ufa}. The data obtained in the following years allowed   measurements of its mass and couplings and a determination of its CP properties and spin \cite{ATLAS:2013sla,CMS:yva,Aad:2013xqa}.  Nevertheless, there are many reasons, both from theoretical considerations and experimental observations, to expect physics beyond the SM, such as the hierarchy problem, neutrino masses and dark matter.  There have been numerous attempts to build new physics models which can explain these puzzles. Some well known examples are the Minimal Supersymmetric Standard Model (MSSM)~\cite{Nilles:1983ge,Haber:1984rc,Barbieri:1987xf}, the Next to Minimal Supersymmetric Standard Model (NMSSM)~\cite{Ellis:1988er,Drees:1988fc} and the Two Higgs Doublet Models (2HDM)~\cite{Branco:2011iw,type1,hallwise,type2}.  

Many of these new physics models involve an extended Higgs sector with an interesting phenomenology that might be testable  at the LHC. In addition to the SM-like Higgs boson in these models, the low energy spectrum includes other CP-even Higgses\footnote{Note that we use $\h$ and $\H$ to refer to the lighter or the heavier CP-even Higgs for models with two CP-even Higgs bosons.  When there is no need to specify, we use $H$ to refer to the CP-even Higgses.} $H$, CP-odd Higgses $A$, and a pair of  charged  Higgses $\hc$. The discovery of one or more of these new particles would be a clear indication of an extended Higgs sector as the source of electroweak symmetry breaking (EWSB). A number of searches have been performed at the LEP, Tevatron and the LHC, mainly focusing on decays of Higgses into SM particles~\cite{LEP_Higgs, Aad:2014vgg, Khachatryan:2014wca, TheATLAScollaboration:2013wia, CMS_taunu, Aad:2013hla, CMS:2014kga, Khachatryan:2015cwa}.   However, exotic decay channels, in which a heavy Higgs decays into  either  two lighter Higgses, or a Higgs plus an SM gauge boson,   open up and can even dominate if kinematically allowed, reducing the reach of the conventional search channels.  Some of these channels have already been studied both in a theoretical~\cite{Curtin:2013fra, Brownson:2013lka, Coleppa:2014hxa, Coleppa:2014cca,Tong_Hpm,Dorsch:2014qja,Chen:2013emb,Chen:2014dma,Enberg:2014pua}  and experimental~\cite{Aad:2015wra, CMS:2014yra,CMS-HZ} setting.  Soon, more of those  exotic Higgs decay channels will   be accessible  at the LHC.   It is therefore timely to study the LHC reach of those channels more carefully.

In the current study we examine the detectability of a light charged Higgs boson, with $m_{H^\pm}<m_t$.  The dominant production mode for such a light charged Higgs at the LHC is via top decay, given the large  top production rate at the LHC.  ${\rm BR}(t\rightarrow H^\pm b)$ can be enhanced at both large and small $\tan\beta$, due to the enhanced top and bottom Yukawa couplings.   Current search strategies assume that the charged Higgs decays either leptonically ($\hc \to \tau\nu$) or hadronically ($\hc \to cs$).  The null search results at both the ATLAS and CMS  exclude a light charged Higgs below a mass of about 160 GeV for most of  the parameter  space~\cite{TheATLAScollaboration:2013wia,CMS_taunu}.  However, if there exists a neutral Higgs ($A/H$) light enough such that the $\hc \to A\w/H\w$ channel is kinematically open, the branching fractions into the conventional final states $\tau\nu$ and $cs$ are suppressed and the exclusion bounds can be significantly weakened.   Due to experimental challenges at low energies, such a light neutral Higgs has not been fully excluded yet.    A relatively large region of $m_{H^\pm}>150$ GeV and $\tan\beta\lesssim 20$ is still allowed, while no limits exist for $m_{H^\pm}>160$ GeV.  

The exotic decay channel of $\hc \to AW/HW$, on the other hand,  offers an additional opportunity for the detection of a light charged Higgs, closing the loophole of the current light charged Higgs searches.  While there are strong constraints on the mass of the light charged Higgs from flavor~\cite{Mahmoudi:2009zx,Coleppa:2013dya} and precision~\cite{EW} observables, those limits are typically model dependent and could be relaxed when there are contributions from the other sectors of the model~\cite{Han:2013mga}.   A direct search for a light charged Higgs, on the other hand,  provides a model-independent  and complementary reach.  It is thus timely and worthwhile to fully explore the discovery or exclusion potential of the light charged Higgs at the LHC.

In this paper we study the exotic decay of a charged Higgs $\hc \to AW/HW$  with $A/H$ decaying into  $\tau\tau$.  We focus on the light charged Higgs  produced via top decay, considering both the single top and top pair production channels. The exclusion bounds and discovery reach will be explored and interpreted in the context of the Type II 2HDM.  A collider analysis considering the same decay channel of a heavy charged Higgs   produced in $\hc tb$ associate production has been performed in \cite{Coleppa:2014cca}.  

We will proceed as follows. In section \ref{sec:motivation}, we   briefly introduce the Type II  2HDM and present scenarios that permit a large branching fraction for the process $\hc \to AW/HW$. In section \ref{sec:limits}, we   summarize the current experimental constraints on a light charged Higgs. In section \ref{sec:analysis},  we present the details of our collider analysis. We investigate the single top and top pair production channels in section \ref{sec:analysis_tj}  and \ref{sec:analysis_tt},   respectively,  and present the model independent 95\% C.L. exclusion and  5$\sigma$ discovery limits for both processes at the 14 TeV LHC with various luminosities in Section \ref{sec:ana_limits}.  In section \ref{sec:implication}, we   discuss the implications of our analysis for the Type II 2HDM and translate our results into reaches in parameter space. We conclude in section \ref{sec:conclusion}.
  

\section{Theoretical Motivation}
\label{sec:motivation}

In the 2HDM,  we introduce two ${ \rm SU}(2)_L$ doublets   $\Phi_{i}$,  $i=1,2$:
 \begin{equation}
\Phi_{i}=\begin{pmatrix} 
  \phi_i^{+}    \\ 
  (v_i+\phi^{0}_i+iG_i)/\sqrt{2}  
\end{pmatrix},
\label{eq:doublet}
\end{equation} 
where $v_1$ and $v_2$ are the vacuum expectation values (vev) of the neutral components which satisfy the relation $\sqrt{v_1^2+v_2^2}=$ 246 GeV after EWSB.  Assuming an additional discrete ${\cal Z}_2$ symmetry imposed on the Lagrangian,  we are left with six free parameters, which can be chosen as    four Higgs masses ($m_{\h}$, $m_{\H}$, $m_A$, $m_{H^{\pm}}$), a mixing angle $\alpha$ between the two CP-even Higgses, and the ratio of the two vacuum expectation values ($\tan\beta=v_2/v_1$).   In the case where a soft breaking of the ${\cal Z}_2$ symmetry is allowed, there is an additional parameter, $m_{12}^2$.    In the Type II 2HDM, $\Phi_1$ couples to the leptons and down type quarks, while $\Phi_2$ couples to the up type quarks.  Details of the Type II 2HDM can be found in the review paper~\cite{Branco:2011iw}.

The Higgs mass eigenstates contain a pair of CP-even Higgses $(\h, \H)$, one CP-odd Higgs $\A$ and  a pair of charged Higgses $H^\pm$, which  can be written as:
\begin{equation}
\left(\begin{array}{c}
\H\\ \h
\end{array}
\right)
=\left(
\begin{array}{cc}
\cos\alpha &\sin\alpha\\
-\sin\alpha&\cos\alpha
\end{array}
\right)  \left(
\begin{array}{c}
\phi_1^0\\\phi_2^0
\end{array}
\right),\ \ \ 
\begin{array}{c}
 A \\H^\pm
 \end{array}
 \begin{array}{l}
 =  -G_1\sin\beta+G_2\cos\beta\\
 =-\phi_1^{\pm}\sin\beta+\phi_2^{\pm} \cos\beta
 \end{array}.
 \label{eq:mass}
 \end{equation}

\begin{figure}[h!]
\centering
	\includegraphics[width=0.45\textwidth]{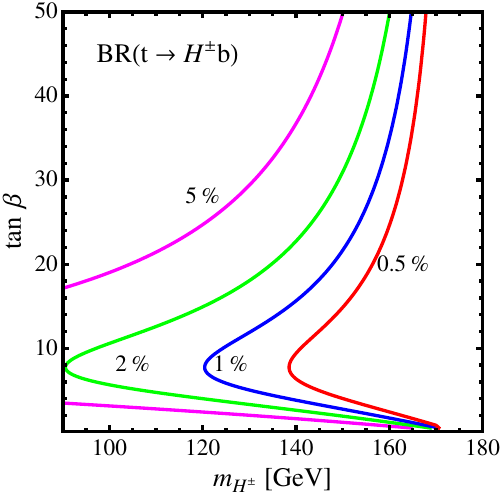}
\caption{Branching fractions of  BR$(t \to \hc b)$ in the $m_{\hc}-\tan\beta$ plane. }
\label{fig:tBR}
\end{figure}

If the charged Higgs is light, the top quark can either decay into $Wb$ or into $H^{\pm}b$. The first decay is controlled by the SM gauge coupling
\begin{equation}
g_{\w tb} = \frac{g}{\sqrt{2}} \gamma^\mu \frac{1-\gamma_5}{2}, 
\end{equation}
with $g$ being the SM ${\rm SU}(2)_L$ coupling, while the second decay depends on  $\tan\beta$ in the Type II 2HDM or MSSM:
\begin{equation}
g_{\hc tb} = \frac{g}{2 \sqrt{2} m_W} \left[ (m_b \tan \beta + m_t \cot \beta ) \pm (m_b \tan \beta - m_t \cot \beta ) \gamma_5 \right].
\end{equation}
This coupling is enhanced for both small and large $\tan\beta$.  In  Fig.~\ref{fig:tBR}, we present contours of the branching fraction BR$(t \to \hc b)$ in the $m_{\hc}-\tan\beta$ plane, calculated  using the 2HDMC \cite{Eriksson:2009ws}.  We can see that the decay branching fraction BR$(t \to \hc b)$ can reach values of 5\% and above for both large and small $\tan\beta$,  but reaches a minimum at $\tan\beta = \sqrt{m_t/m_b}\sim 8 $. The branching fraction decreases rapidly when the charged Higgs mass becomes close to the top mass.

Conventionally, a light charged Higgs is assumed to either decay into $\tau\nu$ or $cs$, with the corresponding couplings being
\begin{eqnarray}
g_{\hc \tau\nu} &=& \frac{g}{2 \sqrt{2} m_W} m_\tau \tan \beta(1\pm    \gamma_5 ), \\
g_{\hc cs} &=& \frac{g}{2 \sqrt{2} m_W} \left[ (m_s \tan \beta + m_c \cot \beta ) \pm (m_s \tan \beta - m_c \cot \beta ) \gamma_5 \right] .
 \label{eqn:hc-smcouplings}
 \end{eqnarray}
If there is an additional light neutral Higgs boson $h^0$ or $A$, additional decay channels into $h^0 W/AW$ open up. The couplings are determined by the gauge coupling structure, as well as the mixing angles \cite{Gunion:1989we}:  
\begin{eqnarray}
 g_{\hc h^0 W^{\mp}}&=& \frac{g\cba}{2}(p_{\h}-p_{\hc})^\mu,  \\
 g_{\hc A W^{\mp}}&=& \frac{g}{2}(p_{A}-p_{\hc})^\mu,  
 \label{eqn:hc-bsmcouplings}
 \end{eqnarray}
 with  $p_\mu$ being the incoming momentum for the corresponding particle. 

The $\hc \to h^0 W$ channel for a light charged Higgs is open only if we demand the heavy CP-even neutral Higgs $\H$  to  be the observed 126 GeV SM-like Higgs. In this case $|\cos(\beta-\alpha)| \sim 1$ is preferred by experiments and the $\hc h^0 \w$ coupling is unsuppressed.  The $\hc A \w$ coupling is independent of $\sba$ and always unsuppressed. There is no   $\hc \to H^0 W$ channel since it is kinematically forbidden given $m_{\hc}<m_t$ and $m_{\H}\geq126$ GeV.

In the generic 2HDM, there are no mass relations between the charged scalars, the scalar and pseudoscalar states. Therefore both the decays $\hc \to h^0 W$ and $\hc \to A W$ can be accessible or even   dominant in certain regions of the parameter space. It was shown in Ref.~\cite{Coleppa:2013dya} that in the Type II 2HDM with ${\cal Z}_2$ symmetry, imposing all experimental and theoretical constraints still leaves large regions in the parameter space that permit such exotic decays with unsuppressed decay branching fractions. 

\begin{figure}[h!]
 \centering
 	\includegraphics[width=0.45\textwidth]{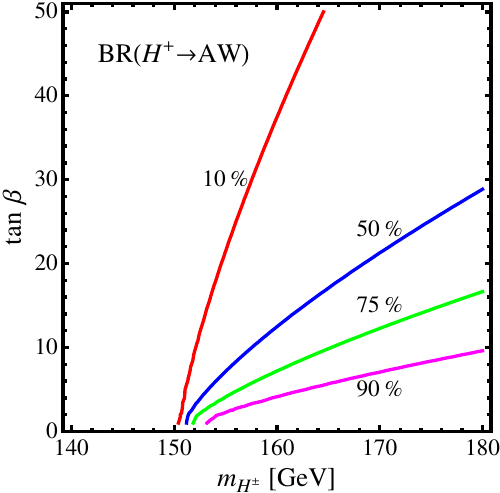}
 	\includegraphics[width=0.466\textwidth,trim=0 +0.05cm 0 0 ]{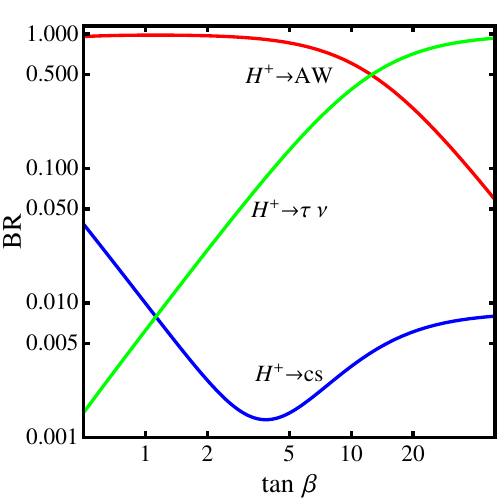}
\caption{The left panel shows the branching fraction BR$(  H^{\pm} \to AW)$ in the Type II 2HDM in $m_{H^\pm}-\tan\beta$ plane. The right panel shows the branching fractions of   $\hc \rightarrow AW$ (red), $\tau\nu$ (green) and $cs$ (blue) as a function of $\tan\beta$ for a 160 GeV $\hc$. Both plots assume the existence of a 70 GeV CP-odd scalar $A$, $h^0$ being the SM-like Higgs and $H^0$ decoupled.}
\label{fig:mod_BR_Hpm}
\end{figure}

In the left panel of Fig.~\ref{fig:mod_BR_Hpm},   we show the contours of the branching fraction BR$(\hc \to A W)$  in the $m_{\hc}-\tan\beta$ plane assuming $m_A=70$ GeV, $h^0$ being the SM-like Higgs and $m_{H^0}$ decoupled. This branching fraction dominates for values of $\tan \beta$ less than 10 to 30 for charged Higgs masses in the range between 155 GeV and 170 GeV. For large values of $\tan\beta$,  the $\tau\nu$ channel dominates, as shown in the right panel of Fig.~\ref{fig:mod_BR_Hpm} for $m_{\hc}=160$ GeV. For small charged Higgs masses close to the $m_A + m_W$ threshold,  the decay is kinematically suppressed. Similar results can be obtained for $\hc \rightarrow \h W$ with $m_{h^0} = 70$ GeV, $\sin(\beta-\alpha)\sim 0$ and decoupled $m_A$. 

The MSSM Higgs mass spectrum is more restricted. At tree level, the mass matrix depends on $m_A$ and $\tan\beta$ only,  and  the charged Higgs mass is related to $m_A$ by $m_{\hc}^2 =m_A^2 + m_W^2$ . Large loop corrections are needed to increase the mass splitting to permit the decay of $\hc \to A W$.  In the non-decoupling region of MSSM with $\H$ being the SM-like Higgs,  the light CP-even Higgs $h^0$ can be light: $m_{h^0}<m_{\hc}-m_W$. The branching fractions can reach values up to 10\% \cite{Heinemeyer:2013tqa} in some regions of parameter space.    In NMSSM the Higgs sector is enlarged by an additional singlet. The authors of \cite{MODEL_NMSSM} have shown that decays of $\hc \to A_iW/H_iW$ can be significant in certain regions of parameter space.

 \section{Current Limits}
 \label{sec:limits}

Searches for a light charged Higgs boson with mass $m_{\hc}<m_t$ have been performed by both ATLAS and CMS~\cite{TheATLAScollaboration:2013wia,CMS_taunu} with 19.7 $\ifb$ integrated luminosity at 8 TeV and 4.6 $\ifb$ integrated luminosity at 7 TeV.  The production mechanism considered is top pair production in which one top quark decays into  $b\hc $ while the other   decays into $bW$. These studies focus on the $\hc \rightarrow \tau \nu$ decay channel, which is dominant in most parts of the parameter space in the absence of decays into lighter Higgses.   Assuming a branching fraction BR$(\hc \rightarrow \tau \nu) = 100 \%$,  the null search results from CMS~\cite{CMS_taunu} imply   upper bounds for the top quark branching fraction BR$(t \rightarrow \hc b)$ varying between   1.2\%  to 0.16\% for charged Higgs masses between 80 GeV and 160 GeV.  This result can be translated into bounds on the MSSM parameter space.  The obtained exclusion limits for the MSSM $m_h^{\rm max}$ scenario can be seen in the right panel of Fig.~\ref{fig:lim_BR_Hpm} (region to the left of red line).   Only charged Higgs masses in the small region 155 GeV $<  m_{\hc} < 160 $ GeV around $\tan \beta = 8$ are still allowed.  The ATLAS results \cite{TheATLAScollaboration:2013wia} are similar.
 
A search with the $\hc \rightarrow cs$ final states has been performed by ATLAS \cite{Aad:2013hla}  using 4.7 fb$^{-1}$ integrated luminosity at 7 TeV and by CMS \cite{CMS:2014kga}  using 19.7 fb$^{-1}$ integrated luminosity at 8 TeV.    Assuming BR$(\hc \rightarrow c s) = 100 \%$,  the ATLAS results imply an upper bound for  BR$(t \rightarrow b \hc )$ around  5\% to  1\% for charged Higgs masses between 90 GeV and 150 GeV while the CMS searches impose an upper bound of BR$(t \rightarrow b \hc ) $  around 2\% to  7\% for a charged Higgs mass between 90 and 160 GeV.

These limits get weaker once we assume realistic branching fractions smaller than 100\%. The left panel of Fig.~\ref{fig:lim_BR_Hpm} shows how the  CMS limits on the branching fraction BR$(t \to \hc b)$ can change  significantly in the presence of an additional light  neutral Higgs.  The black curve shows the CMS limits presented in \cite{CMS_taunu} assuming a 100\% BR$(\hc \to \tau \nu)$.   The modified limits assuming the presence of a 70 GeV CP-odd neutral Higgs are shown for $\tan\beta=1$ (red), $\tan\beta=7$ (blue) and $\tan\beta=50$ (green). We can see that for large $\tan\beta$,  the limits stay almost unchanged since $\hc \to \tau \nu$ is the dominating decay channel, but for smaller values of $\tan\beta$ these limits are weakened significantly.

The right panel of Fig.~\ref{fig:lim_BR_Hpm} shows how the CMS limits in the $m_{\hc}-\tan\beta$ plane  weaken in the presence of an additional light Higgs.   The yellow shaded region (plus the cyan region)  assumes a 100\% BR$(\hc \to \tau \nu)$ while the cyan region assumes the Type II 2HDM branching fractions in the presence of a 70 GeV CP-odd neutral Higgs.     For $\tan\beta<15$,  the surviving region in $m_{\hc}$ is much more relaxed, extending down to about 150 GeV.   Therefore,  the presence of exotic decay modes substantially weakens the current and future limits based on  searches for the conventional $\hc \rightarrow \tau \nu, cs$ decay modes. 

 \begin{figure}[h!]
 \centering
 	\includegraphics[width=0.45\textwidth,]{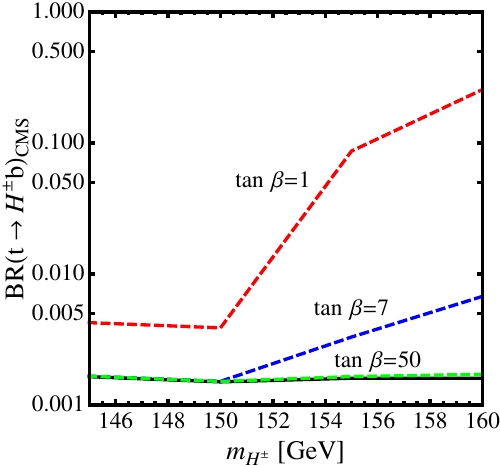}
 	\includegraphics[width=0.408\textwidth,trim=0 -0.1cm 0 0 ]{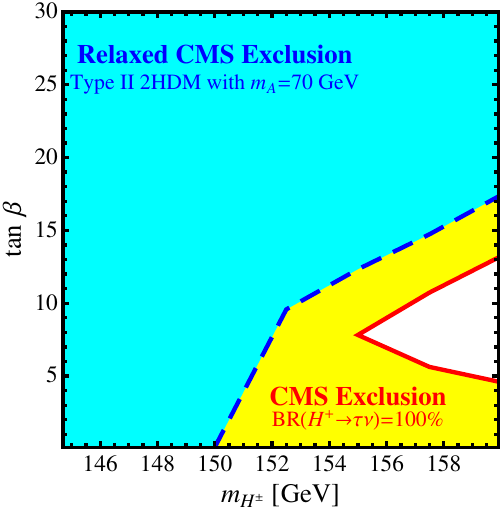}
\caption{Left panel: CMS  limits on the branching fraction BR$(  t \to H^{\pm}b)$ assuming a 100\% BR$(\hc \to \tau \nu)$ (black line)~\cite{CMS_taunu}, as well as the weakened limits in the Type II 2HDM in the presence of a light neutral Higgs for $\tan\beta=1$ (red), $\tan\beta=7$ (blue) and $\tan\beta=50$ (green).  Right panel:  the excluded region in $m_{H^\pm}-\tan\beta$ plane assuming a 100\% BR$(\hc \to \tau \nu)$ (yellow   and cyan regions) and the weakened limits  with a light neutral Higgs (cyan region).  Here we have assumed the light neutral Higgs to be a 70 GeV CP-odd scalar $A$.}
\label{fig:lim_BR_Hpm}
\end{figure}

 A light charged Higgs could have a large impact on precision and flavor observables~\cite{FLAVOR}. For example, in  the 2HDM, the bounds on $b\to s\gamma$ restrict the charged Higgs to be heavier  than 300 GeV. A detailed analysis of precision and flavor bounds in the 2HDM can be found in refs.~\cite{Coleppa:2013dya,Mahmoudi:2009zx}. Flavor constraints on the Higgs sector are, however, typically   model-dependent, and could be alleviated when there are contributions from other new particles in the model~\cite{Han:2013mga}.   Since our focus in this work is   on collider searches for a light  charged Higgs and their implications for the Type II 2HDM, we consider  the scenario of a light charged Higgs: $m_{\hc}<m_t$, as long as it satisfies the direct collider Higgs search bounds.

 Our study also assumes the existence of a light neutral Higgs $A/H$, which has been constrained by the $A/H \rightarrow \tau\tau$ searches at the LHC~\cite{Khachatryan:2014wca,Aad:2014vgg}, in particular, for $m_{A/H}>90$ GeV and relatively large $\tan\beta$.  No limit, however, exists for $m_{A/H}<90$ GeV due to the difficulties in the identification of the relatively soft  taus  and the overwhelming SM backgrounds for  soft leptons and $\tau$-jets.   Furthermore, LEP limits~\cite{LEP_Higgs} based on $VH$ associated production do not apply for the CP-odd $A$ or the non-SM like CP-even Higgs.  LEP limits based on $AH$ pair production can also be avoided as long as $m_A+m_H>208$ GeV.  Therefore, in our analyses below, we choose the daughter (neutral) Higgs mass to be 70 GeV.\footnote{The mass of 70 GeV is also chosen to be above the $h_{\rm SM} \to AA$ threshold to avoid significant deviations of the 126 GeV SM-like Higgs branching fractions from current measurements.}

There have been theoretical studies on other light charged Higgs production and decay channels. The authors of  \cite{THEO_SingleTop} analyzed the possibility of using the single top production mode to observe a light charged Higgs boson decaying into a $\tau\nu$ final state. The detectability of a charged Higgs decay into a $ \mu \nu $ final state  or  a $\gamma\gamma W$ final state via $AW$  with a light charged Higgs produced via top decay in top pair production has been investigated in~\cite{Hashemi:2011gy} and  \cite{Das:2014fha}. 

The $\hc  tb$ associated production with $\hc\to AW/H\w $ has been analyzed in detail in Ref.~\cite{Coleppa:2014cca},  which focuses on heavy charged Higgs bosons ($m_{\hc}>m_t$).  Given the same final state of $bbWWA/H$, the same search strategy can be used to analyze light charged Higgs coming from top decay with top pair production.   Furthermore, we analyze single top production with $pp \to tj$ and $t \to \hc b  \to A/HWb $.   This channel permits a cleaner signal due to its unique kinematic features.

\section{Collider Analysis}
 \label{sec:analysis}

In our analysis we study the exotic decay $\hc \rightarrow AW/HW$ of light charged Higgs bosons $(m_{\hc}<m_t)$ produced via top decay. We consider two production mechanisms: $t$-channel single top production\footnote{We only consider the dominant $t$-channel single top mode since the $s$-channel mode suffers from a very small production rate and the $tW$ mode has a final state similar to that of the  top pair production case.} ($tj$) and top pair production ($t\bar{t}$)~\cite{Kidonakis:2012db}.

 The light neutral Higgs boson can either be the CP-even $H$ or the CP-odd $A$. In the analysis that follows, we use the decay $\hc \rightarrow A \w$ as an illustration. Since we do not  use  angular correlations of the charged Higgs decay,  the bounds obtained for $\hc \rightarrow A \w$ apply to  $\hc \rightarrow H \w$ as well.

The neutral Higgs boson ($A$) itself  will decay further. In this analysis we  look at the fermionic decay $A \rightarrow\tau\tau$ for single top production and both the $\tau\tau$ and the hadronic $bb$  modes  for top pair production.  While the $bb$  mode would have the advantage of a large branching fraction $\text{BR}(A \rightarrow bb)$, the $\tau\tau$ case has smaller SM backgrounds and therefore leads to a cleaner signal. We study both leptonic and hadronic $\tau$ decays and consider  three cases: $\tau_{had}\tau_{had}$,  $\tau_{lep}\tau_{had}$ and  $\tau_{lep}\tau_{lep}$.  The $\tau_{lep}\tau_{had}$ case is particularly promising since we can utilize the same sign dilepton signal with the leptons from the decays of the $W$ and the $\tau$.  

We use Madgraph 5/MadEvent v1.5.11 \cite{Alwall:2011uj} to generate our signal and background events.  These events are passed to Pythia v2.1.21 \cite{PYTHIA}  to simulate initial and final state radiation,  showering and hadronization. The events are further passed through Delphes 3.07 \cite{Favereau:2014} with the Snowmass combined LHC detector card \cite{snowmassdetector} to simulate detector effects. The discovery reach and exclusion bounds have been determined using the program RooStats~\cite{Moneta:2010pm} and theta-auto \cite{thetaauto}.

In this section, we will present model \emph{independent} limits on the $\sigma\times\rm{BR}$ for both 95\% C.L.  exclusion and 5$\sigma$ discovery for both single top and top pair production with possible final states  $\tau\tau bW j$ and $\tau\tau bbWW$/$bbbbWW$.   We consider  the parent particle mass $m_{\hc}$ in the range 150 $-$ 170 GeV and the daughter particle mass,  $m_A=70$ GeV.

\subsection{Single Top Production}
 \label{sec:analysis_tj}
 
For single top production, we consider the channel\
\begin{equation}
pp\to tj\to \hc bj\to AW^{\pm}bj \rightarrow \tau\tau W bj.
\end{equation}    The dominant SM backgrounds are $W\tau\tau$ production, which we generate with up to two additional jets (including $b$ jets); and  top pair production with both fully and semi-leptonic decay chains, which we generate with up to one additional jet. We also take into account the SM backgrounds $tj\tau\tau$ and $ttll$ with $l= (e, \mu, \tau)$.

The cuts that we have imposed are:
\begin{enumerate}

\item \textbf{Identification cuts:} 

\textbf{Case A ($\tau_{had}\tau_{had}$):}  {One lepton $\ell = e$ or $\mu$, two $\tau$ tagged jets, zero or one $b$ tagged jet and at least one untagged jet:} 
\begin{equation}
n_{\ell} = 1, \;  n_{\tau} = 2, \; n_b = 0,1 ,\; n_{j} \geq 1.
\label{eqA}
\end{equation}
We require the $\tau$-tagged jets to have charges of opposite signs.

\textbf{Case B ($\tau_{lep}\tau_{had}$):}  {Two leptons, one $\tau$ tagged jet, zero or one $b$ tagged jet and at least one untagged jet:}
\begin{equation}
n_{\ell} = 2, \; n_{\tau} = 1,\; n_b = 0,1 , \;n_{j} \geq 1.
\label{eqB}
\end{equation}
We require that both leptons have the same sign, which is opposite to the sign of the $\tau$ tagged jet.

\textbf{Case C ($\tau_{lep}\tau_{lep}$):}  {Three leptons, no $\tau$ tagged jet, zero or one $b$ tagged jet and at least one untagged jet:}
\begin{equation}
n_{\ell} = 3, \; n_{\tau} = 0, \; n_b = 0,1 ,\; n_{j} \geq 1.
\label{eqC}
\end{equation}
The following selection cuts for the identification of leptons, $b$ jets and jets are used:
\begin{equation}
|\eta_{\ell,b,\tau}| < 2.5, \; |\eta_{j}| < 5, \; p_{T, \ell_1, j, b} > 20 \text{ GeV and } p_{T, \ell_{2}} > 10 \text{ GeV.}
\label{eqID}
\end{equation}

\item \textbf{Neutrino reconstruction:}  
We reconstruct the  momentum of the neutrino using the missing transverse momentum  and the momentum of the hardest lepton as described in \cite{Aad:2012ux}, assuming that the missing energy is solely from $W\rightarrow \ell \nu$.    In case B and C, the neutrino reconstruction  is relatively poor since there is additional missing energy from the leptonic $\tau$ decay. 

\item \textbf{Neutral Higgs candidate $A$:} The $\tau$ jets (case A), the $\tau$ jet and the softer lepton (case B) or the two softer leptons (case C) are combined to form the neutral Higgs candidate.   {In cases B and C the mass reconstruction is  relatively poor due to missing energy from the neutrino associated with the leptonic $\tau$ decay. }

\item \textbf{Charged Higgs candidate $H^{\pm}$:} The   neutral Higgs candidate, the reconstructed neutrino and the hardest lepton  are combined to form the charged Higgs candidate. 

\item \textbf{Mass cuts:} We place upper limits on the masses of the charged and neutral Higgs candidates, optimized for each mass combination.  For $m_{\hc}=160$ GeV and $m_{A}= 70$ GeV, we impose 
\begin{equation}
m_{\tau\tau} < 48  \text{ GeV and } m_{\tau\tau W} < 148 \text{ GeV.}
\label{eqMASS}
\end{equation}

 \item \textbf{Angular correlation:} A unique kinematical signature of single top production is the distribution of the angle $\theta^{*}$, which is the angle between the top momentum in the $tj$ system's rest frame and the $tj$ system's momentum in the lab frame, as suggested in \cite{Kling:2012up}. The differential distribution for $\cos\theta^{*}$  is shown in the left panel of Fig.~\ref{fig:fig3} for signal (red, solid), $t\bar{t}$ (blue, dotted) and $W\tau\tau$ (green, dotted). The signal tends to peak around $\cos\theta^* \approx -1$ while the background is   flat for $W\tau\tau$ and $t\overline{t}$.\footnote{As shown in \cite{Kling:2012up}, the $\cos\theta^{*}$ distribution for $t\bar{t}$ background would peak around $\cos\theta^{*}=1$ if the top quark could be reliably identified.   However, in this paper we approximate the top quark momentum by the momentum of the charged Higgs candidate, which results in a flat distribution of $\cos\theta^*$ for the $t\bar{t}$ system.  }   In our analysis we require  
\begin{equation}
 \cos\theta^*<  -0.8.
\label{eqT}
\end{equation}

\begin{figure}[htbp]
\centering
\includegraphics[width = 0.49\textwidth]{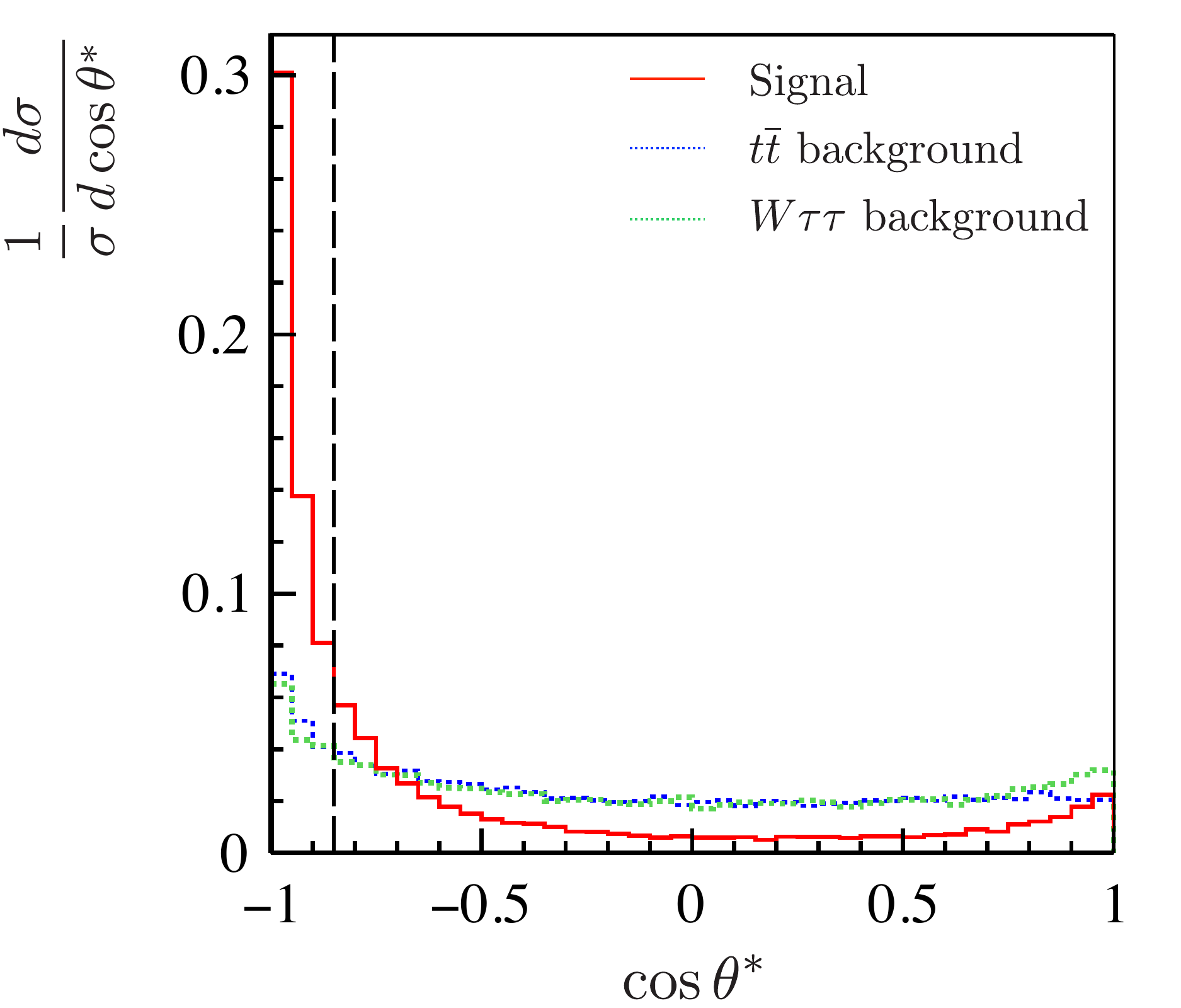}\hfill
 \includegraphics[width = 0.49\textwidth]{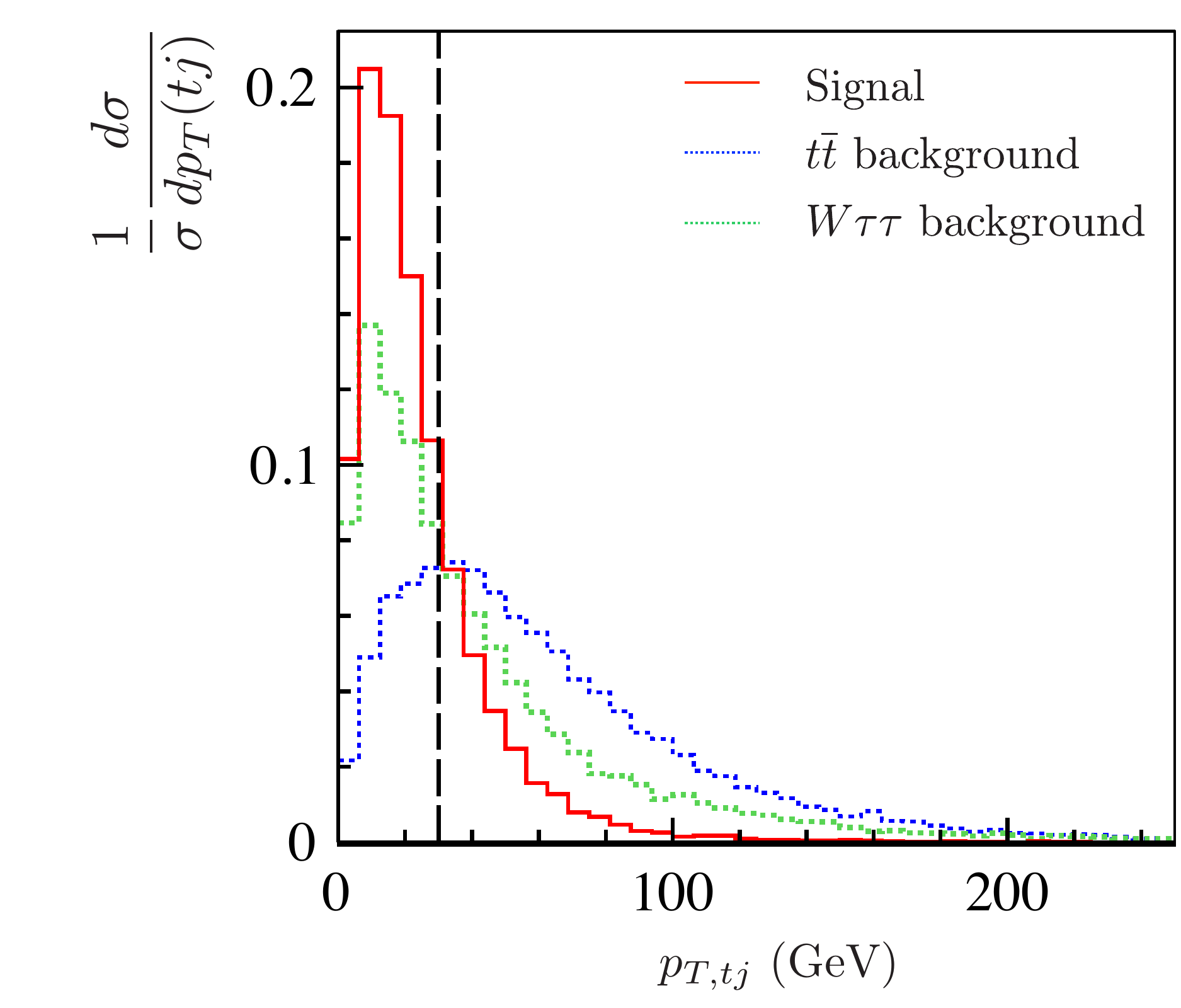}
 \caption{Normalized distribution of $\cos\theta^*$ (left  panel) and the transverse momentum of the $tj$ system $p_{T,tj}$ (right panel) for the signal (red, solid) and the dominant SM backgrounds: $t\bar{t}$ (blue, dotted)  and $W\tau\tau$ (green, dotted). The imposed cuts are indicated by the vertical dashed lines. The histograms shown are for case A with $m_{H^{\pm}}=160\text{ GeV}$ and $m_A=70$ GeV.}
\label{fig:fig3}
\end{figure}

\item\textbf{Top and recoil jet system momentum:} In single top production, we expect that the transverse momentum of the top quark and recoil jet should balance each other, as shown in the right plot of Fig.~\ref{fig:fig3} by the red solid curve.  We impose the cut for the transverse momentum of  the $tj$ system: 
\begin{equation}
p_{T,tj} < 30 \text{ GeV}.
\label{eqPT}
\end{equation}
This further suppresses the top pair background in the presence of additional jets coming from the   second top.
 
\end{enumerate}

In Table \ref{tab:tj}, we show the signal and  major  background cross sections with cuts for a signal benchmark point of $m_{\hc} = 160$ GeV and $m_A =  70$ GeV at the 14 TeV LHC. The first row shows the total cross section before cuts,  calculated using MadGraph.    The following rows show the cross sections after applying the identification cuts, mass cuts and the additional cuts on $\cos\theta^*$ and $p_{T,tj}$ for all three cases as discussed above. We have chosen a nominal value for $\sigma \times \text{BR}( p p  \rightarrow \hc b j \rightarrow \tau \tau W bj)$ of 100 fb.\footnote{For the Type II 2HDM the top branching fraction into a charged Higgs for $m_{\hc}=160$ GeV is typically  between 0.1\% and 1\% (see Fig.~\ref{fig:tBR}). Using the single top production cross section, $\sigma_{tj}=248$ pb \cite{Kidonakis:2012db} and assuming  the branching fractions BR$(\hc \to A\w)  = 100 \%$ and BR$(A \to \tau\tau)=8.6 \%$ leads to the stated $\sigma\times$ BR of around 21 $-$ 210 fb.  }

\begin{table}[h]
\centering
\resizebox{14cm}{!} {
\begin{tabular}{ |l | r |r r r |r r|}  \hline
Cut 		 														&Signal 	&		$W(W)\tau\tau$ 	&	$t\bar{t}$		&$tj\tau\tau/ttll$	&$S/B$	&$S/\sqrt{B}$	\\
																&[fb]		&			[fb]		&	[fb]					&[fb]		&		&	(300 fb$^{-1}$)		\\
\hline
$\sigma$    														&100 		&		  2000		&    $6.3 \cdot 10^5$  				& 257		&	-			&	-	\\
\hline
A: Identification [Eq.(\ref{eqA})] 		 								&0.29	 	&           5.36 		&    130			& 1.39		&	 0.002     	& 0.43	\\
\phantom{A:} 
Mass cuts [Eq.(\ref{eqMASS})]											&0.16	 	&           0.34 		&    2.62				& 0.04		&	 0.05     	& 1.55	\\
\phantom{A:} $\cos\theta^*$ and $p_{T,tj}$ [Eq.(\ref{eqT}), (\ref{eqPT})]&0.07	 	&           0.03 		&    0.07					& 0.001		& 	 0.67     	& 3.72	\\
\hline
B: Identification [Eq.(\ref{eqB})]	 									&0.25	 	&           4.45 		&    2.46					& 1.33	&	 0.03    		& 1.51	\\
\phantom{B:} Mass cuts [Eq.(\ref{eqMASS})]								&0.11	 	&           0.31 		&    0.20					& 0.05	&	 0.19    		& 2.48	\\
\phantom{B:} $\cos\theta^*$ and $p_{T,tj}$ [Eq.(\ref{eqT}), (\ref{eqPT})]&0.06	 	&           0.04 		&    0.02					& 0.002		&	 0.91    		& 3.99	\\
\hline
C: Identification [Eq.(\ref{eqC} )]	 									&0.18	 	&           3.07 		&    6.77				& 6.74		& 	0.01     	& 0.78  \\
\phantom{C:} Mass cuts [Eq.(\ref{eqMASS})]								&0.12	 	&           0.55 		&    0.94				& 0.28		& 	0.07     	& 1.63  \\
\phantom{C:} $\cos\theta^*$ and $p_{T,tj}$ [Eq.(\ref{eqT}), (\ref{eqPT})]&0.07	 	&           0.08 		&    0.10					& 0.01		& 	0.38     	& 2.84 \\
\hline
\end{tabular}
}
\caption{Signal and dominant background cross sections with cuts for the signal benchmark point $m_{\hc}$ = 160 GeV   and $m_A$ = 70 GeV at the 14 TeV LHC.  We have chosen a nominal value for $\sigma \times {\rm BR}(pp \rightarrow tj \to \hc  j b \rightarrow \tau\tau W b j )$ of 100 fb to illustrate the cut efficiencies for the signal process.  The last column of $S/\sqrt{B}$ is shown for an integrated luminosity of ${\cal L}=300\  {\rm fb}^{-1}$.  }
\label{tab:tj}
\end{table}

We can see that the dominant background contributions after particle identification are $t\overline{t}$ for cases A and C, and $W\tau\tau$   for case B.  The reach is slightly better in case B in which the same sign dilepton signature can reduce the $t\bar{t}$ background sufficiently.  Nevertheless, soft leptons from underlying events or $b$-decay can mimic the same sign dilepton signal.  The obtained   results are sensitive to the $\tau$ tagging efficiency as well as the misidentification rate.  In our analyses, we have used a  $\tau$ tagging efficiency of $\epsilon_{tag}=60\%$ and a mistagging rate of $\epsilon_{miss}=0.4\%$,  as suggested in \cite{snowmassdetector}.  A better rejection of non-$\tau$ initiated jets would increase the significance of this channel. 

\subsection{Top Pair Production }
 \label{sec:analysis_tt}

We now turn to the top pair production channel 
\begin{equation}
pp \rightarrow tt \to \hc tb \rightarrow  AbbWW\rightarrow  \tau\tau bbWW/bbbbWW.
\end{equation} 
A detailed collider study with the same final states has been performed in \cite{Coleppa:2014cca} with a focus on high charged Higgs masses. The same strategy has been adopted for the light charged Higgs case and we refer to Ref.~\cite{Coleppa:2014cca} for details of the analysis. 

To analyze this channel,  we consider   decay modes of the neutral Higgs into $\tau_{had}\tau_{had}$, $\tau_{had}\tau_{lep}$, $\tau_{lep}\tau_{lep}$ and $bb$.    For the two $W$ bosons, we require one to decay leptonically and the other to decay hadronically to reduce backgrounds.
  
 The dominant SM background for the $\tau\tau$ channel is semi- and fully leptonic $t\bar{t}$ pair production. We also take into account  $ttll$ production with $l = (e, \mu, \tau)$, as well as $W\tau\tau$ and $WW\tau\tau$.  We ignored the subdominant backgrounds from single vector boson production, $WW$, $ZZ$, single top production, as well as multijet QCD background.   Those backgrounds are either small or can be sufficiently suppressed by the cuts imposed.   Similar backgrounds are considered for the $bb$ process.   

In Table \ref{tab:tt}, we show the signal and  major  background cross sections of the $\tau\tau$ channel with cuts for a signal benchmark point of $m_{\hc} = 160$ GeV and $m_A =  70$ GeV at the 14 TeV LHC, similar to Table \ref{tab:tj}.   We have chosen a nominal value for $\sigma \times {\rm BR}(pp \rightarrow tt \to \hc tb \rightarrow  \tau\tau bb WW)$ of 1000 fb to illustrate the cut efficiencies for the signal process. 

After the cuts,   the dominant background contributions are $t\bar{t}$ ($\tau_{had}\tau_{had}$,  $\tau_{lep}\tau_{lep}$) as well as  $t\bar{t}ll$ ($\tau_{had}\tau_{lep}$)    while the backgrounds including vector bosons do not contribute much. We find  that the case in which one $\tau$ decays leptonically and the other $\tau$ decays hadronically gives the best reach. This is because the same sign dilepton signature can reduce the $t\bar{t}$ background sufficiently.

\begin{table}[h]
\centering
\resizebox{14cm}{!} {
\begin{tabular}{ |c | r |r r r |r r|}  \hline
Cut 		 								&Signal [fb] 	&$t\bar{t}$ [fb] 		&$t\bar{t}ll$ [fb]	&$W(W)\tau\tau$ [fb]	&$S/B$	&$S/\sqrt{B}$	\\
\hline
$\sigma$    								&1000 		&$6.3 \cdot 10^5$	&  247   		   	& 2000    		 &		 	 &		 \\
\hline
$\tau_{had}\tau_{had}$: Identification  		 			&4.1	 		&           23.3 		&    0.58			 			& 0.078     		& 0.17	&	14.9		\\
\ \  \ $m_{\tau\tau}$ vs $m_{\tau\tau W}$  			&0.6   		&           0.31 		&    0.021 			 			& 0.003     		& 1.9  	&	18.8		\\
\hline
 $\tau_{lep}\tau_{had}$: Identification 	 				&3.3 			&           0.35 		&    0.697 			 			& 0.072     		& 3.0		&	55.3		\\
 \ \ \  $m_{\tau\tau}$ vs $m_{\tau\tau W}$ 			&0.69 		&           0.035 		&    0.042 			 			& 0.007   		& 8.1		&	41.1		\\
\hline
 $\tau_{lep}\tau_{lep}$: Identification  	 				&3.1			&           2.35 		&    5.11 			 				& 0.058     		& 0.41	&	19.9		\\
 \ \  \ $m_{\tau\tau}$ vs $m_{\tau\tau W}$ 			&0.62 		&           0.25 		&    0.16 			 				& 0.006         	& 1.4		&	16.5		\\
 \hline
\end{tabular}
}
\caption{Signal and background cross sections with cuts for the signal benchmark point $m_{\hc}$ = 160 GeV   and $m_A$ = 70 GeV at the 14 TeV LHC.  We have chosen a nominal value for $\sigma \times {\rm BR}(pp \rightarrow tt \to \hc tb \rightarrow   \tau\tau bb WW)$ of 1000 fb to illustrate the cut efficiencies for the signal process.  The last column of $S/\sqrt{B}$ is shown for an integrated luminosity of ${\cal L}=300\  {\rm fb}^{-1}$.   See details in Ref.~\cite{Coleppa:2014cca} for the identification cuts and  $m_{\tau\tau}$ vs $m_{\tau\tau W}$ cuts.        }
\label{tab:tt}
\end{table}

\subsection{Limits}
\label{sec:ana_limits}

Fig.~\ref{fig:ana_limits} displays the  95\% C.L. exclusion (green curve) and 5$\sigma$ discovery (red curve) limits at the 14 TeV LHC for both the single top (left) and top pair (right) channel . The dot-dashed, solid and dashed line show the results for three luminosities: 100 fb$^{-1}$, 300 fb$^{-1}$  and 1000 fb$^{-1}$,  respectively. In these plots we have combined all three cases of $\tau$ decays. While in the single top channel,  all three cases contribute roughly the same to the overall significance,  the highest sensitivity in the top pair production channel  comes from the $\tau_{lep}\tau_{had}$ case. Due to the small number of events in both channels, the statistical error dominates over the assumed 10\% systematic error in the background cross sections.    Therefore, higher luminosities lead to  better reaches.    Assuming 300 ${\rm fb}^{-1}$   integrated luminosity, the 95\% C.L. limits on $\sigma\times{\rm BR}$ are about 35 and 55 fb for the single top and top pair production processes respectively.   The discovery reaches are about 3 times higher. 

Assuming a 100\% branching fraction BR$(\hc \to A W )$ and  BR$(A \to \tau\tau)=8.6\%$\footnote{Assuming $bb$ and $\tau\tau$ are the dominant decay modes of a light $A$, BR$(A \to \tau\tau)=8.6$\% in the Type II 2HDM or MSSM at medium to large $\tan\beta$. This branching fraction decreases for small $\tan\beta$ when the $cs$-channel is enhanced.},  we can reinterpret $\sigma\times{\rm BR}$ limits as limits on the branching fraction ${\rm BR}(t \to \hc b)$ as indicated by the vertical axis on the right.   While the cross section limits are better in the single top channel,   the corresponding limits on the branching fraction ${\rm BR}(t \to \hc b)$ are weaker due to the smaller single top production cross section.   The 95\% C.L. exclusion limit on BR$(t \to \hc b)$ is about 0.2\% for the single top process and  0.03\% for the top pair production process, respectively.

A study of the $A \to bb$ decay using the top pair production channel leads to worse results due to the significantly higher SM backgrounds.  For the 14 TeV LHC with 300 fb$^{-1}$, the exclusion limit on $\sigma \times {\rm BR}$ is about 7 pb for a charged Higgs with mass $m_{\hc} = 160$ GeV,  assuming the existence of a light neutral Higgs with mass $m_A=70$ GeV. Thus, given the typical ratio of ${\rm BR}(A/H \to bb) : {\rm Br}(A/H\to \tau\tau) \sim 3m^2_b/m^2_{\tau}$, we conclude that the reach in the $bb$ case is much worse than that in  the $\tau\tau$ case.

 \begin{figure}[h!]
 \centering
 	\includegraphics[width=0.45\textwidth]{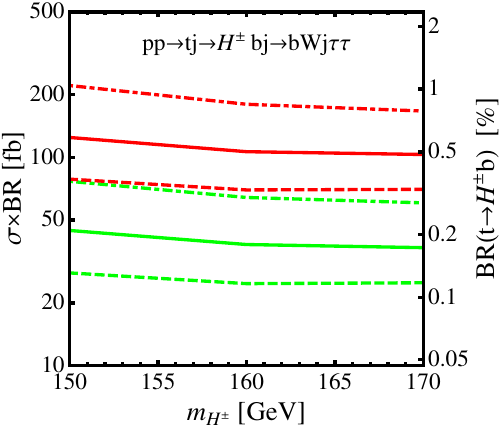} \hspace{0.2 in}
 	\includegraphics[width=0.45\textwidth]{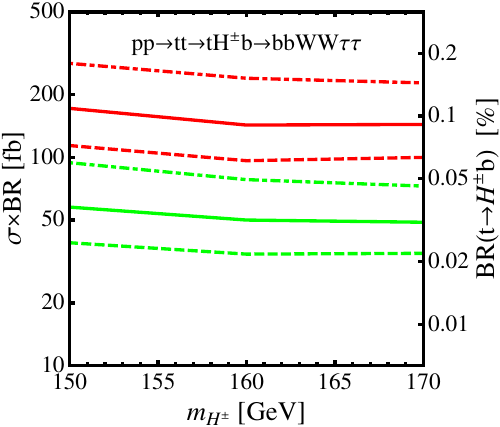}
\caption{The 95\% C.L. exclusion (green) and 5$\sigma$ discovery (red) limits for $\sigma \times $BR and BR$(t \to \hc b)$ (right vertical axis) assuming  BR$(\hc  \to A W)=100\%$ and  BR$(A \to \tau\tau)=8.6\%$ for $m_A = 70$ GeV at the 14 TeV LHC using the single top (left panel) and top pair (right panel) production channels. The dot-dashed, solid and dashed lines correspond to an integrated luminosity of 100, 300 and 1000 fb$^{-1}$ respectively. Here, we have assumed a 10\% systematic error on the backgrounds.   } 
\label{fig:ana_limits}
\end{figure}

We reiterate here that the exclusion and discovery limits  on $\sigma \times {\rm BR}$  are completely model independent. Whether or not discovery/exclusion is actually feasible in this channel should be answered within the context of a particular model, in which the theoretically predicted cross sections and branching fractions can be compared with the exclusion or discovery limits.   We will do this in Sec.~\ref{sec:implication} using the Type II 2HDM as a specific example. 

\section{Implication for the Type II 2HDM}
 \label{sec:implication}

 The results in the previous section on ${\rm BR}(t \rightarrow b \hc)$ can be applied to any beyond the SM scenarios containing   a light charged Higgs boson with the $\hc \to AW/HW$ channel being kinematically accessible. To give a specific example of the   implication of this channel, we will now apply the exclusion and discovery limits in the context of the Type II 2HDM. 

 The 2HDM allows us to interpret the observed Higgs signal either as the lighter CP-even Higgs ($h^0$-126) or the heavier CP-even Higgs ($H^0$-126). The authors of Ref.~\cite{Coleppa:2013dya}  have identified  the Type II 2HDM parameter space in both cases, assuming  $m_{12}^2=0$  and including all the experimental and theoretical constraints. In the $h^0$-126 case,  we are restricted to either a SM-like region at $\sin(\beta-\alpha)=\pm1$ with $\tan\beta<4$ or an extended region with $0.6<\sin(\beta-\alpha)<0.9$ and $1.5<\tan\beta<4$ with relatively unconstrained masses. In the $H^0$-126 case,  a SM-like region, around $\sin(\beta-\alpha)=0$ and $\tan\beta<8$, and an extended region with $-0.8 < \sin(\beta-\alpha)<0.05$ and $\tan\beta$ up to 30 or higher, survive all constraints.   
 
We can interpret the results of the previous section in two ways: the light neutral Higgs in the charged Higgs decay could either be the light CP-even Higgs $h^0$ or the CP-odd Higgs $A$.   The decay mode $\hc \to H^0 W$ is not possible given that $m_{\H}\geq 126$ GeV.   The decay $\hc \to AW$ is  possible in both the $h^0$-126 and $H^0$-126 case and the partial decay width is independent of $\sin(\beta-\alpha)$.   The decay branching fraction, however, depends on whether $\hc \to \h W$ is open or not.  For simplicity, we choose  a benchmark point BP1, with $\left\{ {m_{\hc},m_{\A},m_{\h},m_{\H}}\right\}=\left\{ {160, 70, 126, 700}\right\}$ such that only $\hc \to AW$ is kinematically accessible.   The decay width $\hc \to h^0 W$ depends on $\sin(\beta-\alpha)$ and is only sizable in the $H^0$-126 case.  We illustrate this case with  a second benchmark point  BP2: $\left\{ {m_{\hc},m_{\A},m_{\h},m_{\H}}\right\}=\left\{ {160, 700, 70, 126}\right\}$, assuming that the CP-odd Higgs $A$ decouples.  We list the benchmark points  in Table~\ref{tab:classification}. 

 \begin{table}[h]
\begin{center}
  \begin{tabular}{|l|c|c|c| l | }
    \hline
    $\left\{ {m_{\hc},m_{\A},m_{\h},m_{\H}}\right\}$ GeV & $\hc\to\A W$ & $\hc\to\h W$ & Favored Region  \\ \hline
        BP1: $\left\{ {160, 70, 126, 700}\right\}$ & \cmark & \xmark & $\sba\approx\pm$ 1 \\ \hline
   BP2: $\left\{ {160, 700, 70, 126}\right\}$ & \xmark & \cmark &  $\sba\approx $ 0 \\ \hline
   \end{tabular}
\end{center}
\caption{Benchmark points used for illustrating the discovery and exclusion limits in the context of the Type II 2HDM. The checkmarks indicate kinematically allowed channels. Also shown are the typical favored region of $\sin(\beta-\alpha)$  for each case (see Ref.~\cite{Coleppa:2013dya}). }  
\label{tab:classification}
\end{table}

In the left panel of Fig.~\ref{fig:imp_BP_BR},  we show the branching faction  BR$(\hc \to AW)$ for BP1, which is independent of $\sin(\beta-\alpha)$ and decreases with increasing $\tan\beta$ due to the enhancement of the $\tau\nu$ mode.  The branching fraction can reach values of 90\% or larger for small $\tan\beta<4$ and stays the dominating channel until $\tan\beta=12$.

 \begin{figure}[h!]
 \centering
 	\includegraphics[width=0.45\textwidth]{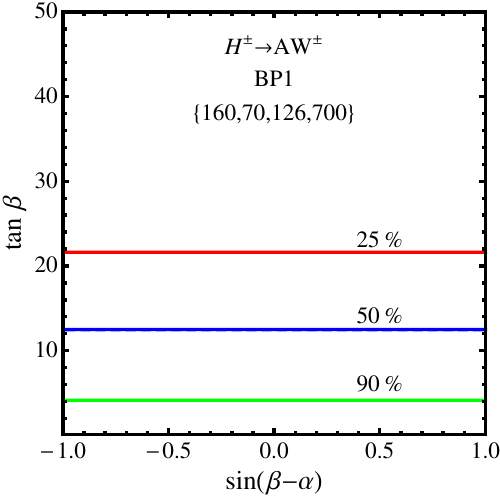}
	\includegraphics[width=0.45\textwidth]{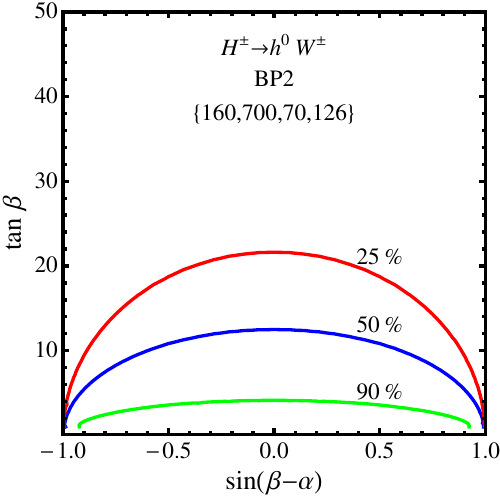}
\caption{Contours of branching fractions of $\hc \to AW$ (left  panel) and $ \hc \to \h W$ (right  panel) for BP1 and BP2,  respectively. }
\label{fig:imp_BP_BR}
\end{figure}

The right panel of Fig.~\ref{fig:imp_BP_BR} shows the branching fraction, BR$(\hc \to h^0W)$, for BP2. It reaches maximal  values around $\sin(\beta-\alpha)=0$ and decreases for larger $|\sin(\beta-\alpha)|$ compared to BP1 due to the suppressed $\hc h^0 W$ coupling. 

In Fig.~\ref{fig:imp_BP_Lim}, we display the 95\% exclusion (yellow regions enclosed by the solid lines as well as the cyan regions)  and 5$\sigma$ discovery reach (cyan regions enclosed by the dashed lines) for BP1 (left panel)  and BP2 (right panel) at the 14 TeV LHC with 300 ${\rm fb}^{-1}$ integrated luminosity.    The red lines refer to the limits based on top pair production, and the blue lines refer to the limits based on single top production.     

 For  the benchmark point  BP1 with $\hc \rightarrow \A\w$, the exclusion reach based on top pair production  covers the entire parameter space, while   discovery   is  possible for small $\tan\beta<6$ and  large $\tan\beta>18$,  independent of $\sba$. Intermediate values of $\tan\beta$ have a reduced branching fraction BR$(t \to \hc b)$ (see Fig.~\ref{fig:tBR}) and therefore the total $\sigma \times$BR is suppressed. At high $\tan\beta$,   BR$(t \to \hc b)$ is enhanced sufficiently to overcome the reduced branching fraction BR$(\hc \rightarrow  AW)$.  The search based on single top production is only effective in the small $\tan\beta$ region, with an exclusion reach of $\tan\beta <4$ and a discovery reach of $\tan\beta<2$.   

 \begin{figure}[h!]
 \centering
 	\includegraphics[width=0.45\textwidth]{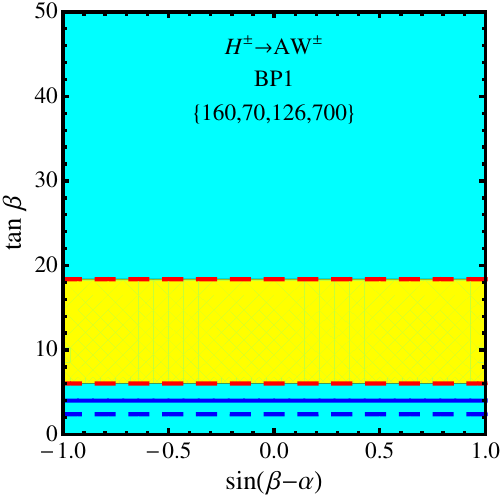}
	\includegraphics[width=0.45\textwidth]{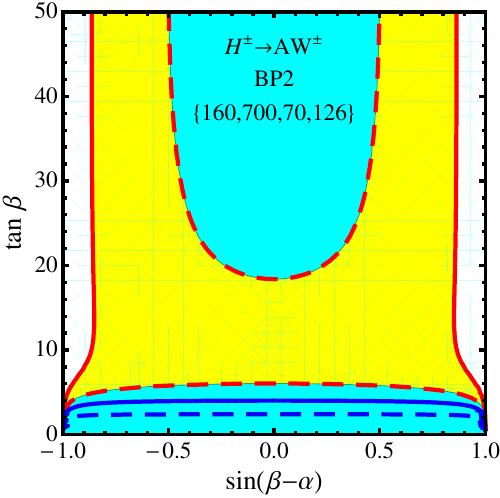}
\caption{The 95\% exclusion (yellow regions enclosed by the solid lines as well as the cyan regions) and the $5\sigma$ discovery reach (cyan regions enclosed by the dashed lines) obtained by the $tj$-channel (blue) and $tt$-channel (red) in the $\tan\beta$ versus $\sba$  plane for BP1 (left panel) and BP2 (right panel), with an integrated luminosity of 300 fb$^{-1}$ at the 14 TeV LHC.  }
\label{fig:imp_BP_Lim}
\end{figure}

The right panel of Fig.~\ref{fig:imp_BP_Lim} shows the reach for BP2. The exclusion region for top pair production covers the entire parameter space except for $|\sin(\beta-\alpha)| >0.85$ and $\tan\beta> 4$.    Discovery is   possible for large $\tan\beta>18$ with  $|\sin(\beta-\alpha)|<0.5$ and for small $\tan\beta<6$.    The reach for single top production is limited to the small $\tan\beta$ region.

In Fig.~\ref{fig:imp_Limits},  we show the reach in the $m_{\hc}-\tan\beta$ plane for $\hc \to AW$ with $m_A=70$ GeV with both $\h$ and $\H$  outside the kinematic reach.   These limits also apply for $\hc \to h^0W$ with $m_{h^0}=70$ GeV and $\sin(\beta-\alpha)=0$ with a decoupled $A$. We display the 95\% exclusion (yellow regions enclosed by the solid lines as well as the cyan regions)  and 5$\sigma$ discovery limits (cyan regions enclosed by the dashed lines)  for an integrated luminosity of 300 fb$^{-1}$ at the 14 TeV LHC. Superimposed are the current CMS limits (black hatched region)~\cite{CMS_taunu} which exclude the large  $\tan\beta$ region at $m_{H^\pm}<160$ GeV. 

The best reach is obtained by the top pair channel, as indicated by regions enclosed by the red lines. The model can be excluded up to 167 GeV for all $\tan\beta$ and up to 170 GeV for $\tan\beta<4$ or $\tan\beta>29$. Discovery is possible for both low $\tan \beta<$6  in the entire region of  150 GeV $< m_{H^\pm}<170$ GeV and high $\tan\beta > 17$ with  155 GeV $< m_{H^\pm}<165$ GeV.    The reach is weakened for intermediate $\tan\beta$ due to the reduced branching fraction $t \to  \hc b$.   The single top channel (blue lines) only provides sensitivity in the low $\tan\beta$ region and permits exclusion (discovery) for $\tan\beta \lesssim$ 4 (3).

 \begin{figure}[h!]
 \centering
 	\includegraphics[scale=0.8]{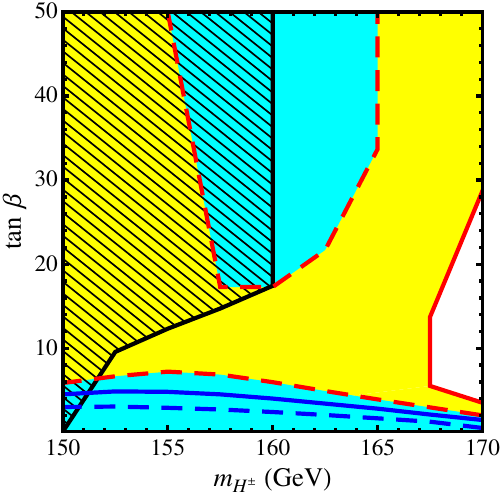}
\caption{95\% exclusion (yellow regions bounded by solid lines as well as the cyan regions) and the $5\sigma$ discovery (cyan regions bounded by the dashed lines) imposed by the $tj$-channel (blue) and $tt$-channel (red) in the $m_{\hc}-\tan\beta$ parameter space for 300 fb$^{-1}$ luminosity with $m_\A=$ 70 GeV. The same limits apply for $m_{h^0}=70$ GeV and $\sin(\beta-\alpha)=0$ if $A$ is decoupled. The black hatched region indicates the region excluded by the CMS search based on $\hc \rightarrow \tau \nu$~\cite{CMS_taunu}. }
\label{fig:imp_Limits}
\end{figure}

 We conclude this section with the following observations:  
\begin{itemize}
\item Once the $AW/h^0W$ channels are kinematically accessible, they dominate for small and intermediate values of $\tan\beta$. The reach in the $\hc \to \tau\nu$ mode is significantly weakened in the presence of the $\hc \to AW/h^0W$ modes, in particular for small  to intermediate $\tan\beta$, leaving the possibility of a light charged Higgs that has escaped detection so far.
\item Both the $\hc \to A W$ channel for the $\h$-126 case and the $\hc \to \h W$ channel in the $\H$-126 case permit exclusion and discovery in large regions of the parameter space.
\item The reach in the exotic channels  $\hc \rightarrow \A W/h^0 W$ is complementary to the reach in the conventional search channel  $\hc \to \tau\nu$,  especially for small to intermediate values of $\tan\beta$.   
\item While the top pair production channel covers a large  region of parameter space, the single top channel permits discovery/exclusion  in the  low $\tan\beta$ region.      
\end{itemize} 

\section{Conclusion}
\label{sec:conclusion}

After the discovery of the first fundamental scalar by both the ATLAS and CMS collaboration, it is now time to carefully measure its properties to determine the nature of this particle. Current measurements still permit the possibility that the discovered signal is not the SM Higgs particle, but just one scalar particle contained in a larger Higgs sector, as predicted by many  extensions of the SM. While most of the current searches for the non-SM  Higgs bosons focus on conventional search channels,  increasing attention is being paid to exotic Higgs decay channels \cite{Curtin:2013fra, Brownson:2013lka, Coleppa:2014hxa, Coleppa:2014cca,Tong_Hpm,Dorsch:2014qja,Chen:2013emb,Chen:2014dma,Enberg:2014pua,CMS:2014yra,Aad:2015wra,CMS-HZ}  into a pair of lighter Higgses or a Higgs plus vector boson final states that can become dominant once kinematically allowed.

In this paper we consider the possibility of a light charged Higgs $m_{\hc} < m_t$ produced via top decay $t \to \hc b$. Due to the large single top and top pair production cross section at the LHC, the charged Higgs can be  produced copiously. Assuming that a light charged Higgs predominantly decays into $\tau\nu$, both ATLAS and CMS exclude a light charged Higgs for most  regions  of the  MSSM and the Type II 2HDM  parameter spaces. The branching fraction BR$(\hc \to \tau\nu)$ can be significantly reduced once the exotic decay channel into a light Higgs, $\hc \to AW/HW$, is open.   In this case, the exclusion bounds from the $\tau\nu$ search get weakened,   in particular for small and intermediate $\tan \beta$,   leaving the possibility of a light charged Higgs open.    This loophole, however, can be closed when we consider the alternative charged Higgs decay channel: $\hc \to AW/HW$.

In this paper we analyze the possibility of discovering a light charged Higgs via the $\hc \to AW/HW$ decay mode assuming that the light Higgs $A/H$ decays into either $\tau\tau$ or $bb$. While the top pair channel benefits from a large production cross section, the single top channel permits a cleaner signal due to its unique kinematic features. Assuming the existence of a light neutral Higgs of mass 70 GeV, the model independent  95\% C.L. exclusion limits on $\sigma\times$BR based on $\tau\tau$ channel  are about 35 fb for the single top channel and 55 fb for the top pair channel. The discovery reaches are about three times higher.   Assuming ${\rm BR}(\hc \rightarrow AW/HW)=100\%$ and ${\rm BR}(A/H \rightarrow \tau\tau)=8.6\%$,  the exclusion limits on ${\rm BR}(t \rightarrow H^+ b)$ are about 0.2\% and 0.03\% for single top and top pair production, respectively.  A significantly worse reach is obtained in the $bb$ channel. 

We discuss the implications of the obtained exclusion and discovery bounds in the context of the Type II 2HDM, focusing on two scenarios: the decay $\hc \to AW$ with a light $A$ in the $\h$-126 case and the decay $\hc \to \h W$ in the $\H$-126 case.     The top pair channel provides the best reach and permits discovery for both large $\tan\beta>17$ around $m_{\hc}=160$ GeV and small $\tan\beta<6$ over the entire mass range, while exclusion is possible in the entire $\tan\beta$ versus $m_{\hc}$ plane except for charged Higgs masses close to the top threshold. The single top channel is sensitive in the low $\tan\beta$ region and permits discovery for $\tan\beta< 3$. In particular, the low $\tan\beta$ region is not constrained by searches in $\tau\nu$ channel, making the $\hc \to AW/\h W$ a complementary channel for charged Higgs searches.

While most of the recent searches for additional Higgs bosons have focused on conventional decay channels, searches using exotic decay channels have just started~\cite{Curtin:2013fra, Brownson:2013lka, Coleppa:2014hxa, Coleppa:2014cca,Tong_Hpm,Dorsch:2014qja,Chen:2013emb,Chen:2014dma,Enberg:2014pua,CMS:2014yra,Aad:2015wra,CMS-HZ}.   Studying all of the possibilities for the non-SM Higgs decays will allow us to explore the full potential of the LHC and future colliders in  understanding the nature of electroweak symmetry breaking.

\acknowledgments

We thank B. Coleppa for his participation at the beginning of this project.  We would also like to thank Peter Loch and Matt Leone for helpful discussions. This work was supported by the Department of Energy under Grant~DE-FG02-13ER41976.

 \bibliographystyle{JHEP}

\end{document}